# Electronic and magnetic properties of new binary uranium-boron compounds with 2D and 3D boron networks: A revisit of the U:B system.

Samir F. Matar*

Lebanese German University. LGU. Sahel-Alma, Jounieh P.O.B. 206. Lebanon.

*Formerly at the University of Bordeaux, ICMCB–CNRS). Pessac. France.

email: s.matar@lgu.edu.lb and abouliess@gmail.com

## Abstract

*Based on crystal chemistry rationale and calculations within the density functional theory DFT, the U:B system is complemented with additional binary compounds $UB_3$, $U_2B_6$, and $UB_6$ possessing two-dimensional 2D and 3D boron substructures. Observations are supported quantitatively with the trends of cohesive energies, charge transfers onto the boron sub-lattice and geometry optimized structures. The results point out to a 'structure crossover' from hexagonal (layer B network) to 3D boron network at compositions above $UB_3$ found to be connected with a threshold amount of charge onto boron which is ~0.46. From the energy-volume of states EOS considering spin degenerate and spin-polarized configurations, hexagonal $UB_3$, and cubic $UB_6$ were found in a stable ferromagnetic ground state with 1.47 $\mu_B$ and 2.40 $\mu_B$ spin-only moments. The volume variations of magnetization show respectively a smooth and abrupt evolution for $UB_3$ and $UB_6$.*

# 1 Introduction

The existence and stability of chemical compounds arise from the exchange of electrons between the constituents, establishing the so-called iono-covalent bond. This represents a global albeit vague picture between two extremes: the ionic bond involving a complete transfer from the cation ($Na^+$) to the anion ($F^-$) as in NaF on one hand, and on the other hand a representation of the situation where electrons are in-between the atoms in the covalent bond observed in other compounds as diamond and intermetallic alloys. The criteria allowing a more rationalized approach are the electronegativity ($\chi$), as well as the chemical hardness $\eta$ (cf. [1] for a review).

However another aspect relevant to dimensionality of the structure may be added to the electron transfer criterion; we illustrate it in this introduction with chemical systems based on the early *p* elements, B and C. As it is well established, with carbon, tetrahedral coordination ($sp^3$) –three-dimensional 3D– is found in diamond, making it a perfectly covalent chemical system with the involvement of all 4 valence electrons in $\sigma$ strong bonds within *C4* tetrahedra thus providing electronic structure description for its extreme hardness (light small elements, short distances, and covalent C-C bonds). Oppositely, planar C with $sp^2$ hybridization –two-dimensional 2D– prevails in soft graphite with three in-plane electrons forming $\sigma$ bonds within elementary triangular *C3* motifs and one off-plane $\pi$ electron. However such 2D–configuration is not stable for carbon neighbor, boron which has one electron less. Nevertheless, planar B networks exist: in fact, they are stabilized through the acquisition by B of the necessary charge through the bonding with electropositive elements: M→B. This is found in binary borides as $AlB_2$-type (space group *P6/mmm*) and other isostructural binary compounds as with alkaline-earth M= Mg [2], actinide M= U [3], transition metal M= Cr [4], etc. It needs to be stressed here that in these structures –cf. Fig. 1a– the boron hexagonal substructure is limited to *B2* and the next boron-rich compositions start at $MB_4$ then $MB_6$ which are three-dimensional structures (3D). To address this point, an extension of the hexagonal boron network to *B6* within the same 2D space group as $AlB_2$ –type (*P6/mmm*) was recently proposed in Cr–B system with the compositions: $Cr_2B_6$ and $CrB_6$, both possessing a 2D boron network [5]. The investigations were carried out based on quantum mechanics calculations in the framework of the density functional theory DFT [6,7] which is equally used in present work.

Back in 2000, we investigated the electronic structures and the bonding properties of UB$_2$, UB$_4$, and UB$_{12}$ [8] using *ad hoc* experimental parameters and including hypothetic "UB$_6$" in cubic CaB$_6$ –type structure with an average lattice constant based on ThB$_6$ to allow investigations comparatively within the series [9]. The analyses of the bonding properties and the detailed electronic density of states were provided; however, neither cohesive energy nor geometry minimized crystal results, especially for the devised UB$_6$.

The present paper revisits the U:B system and reports on DFT results of electronic and magnetic structure investigations of binary borides, letting establish trends of cohesive energies as well as energy derived properties as the energy-volume equation of state. Specifically, new uranium compounds with extensions to *B3* and *B6* sublattices are presented as cohesive and possessing a magnetic order for ground state hexagonal UB$_3$ and cubic UB$_6$ versus less stable cubic and hexagonal structures respectively. The crossover from 2D to 3D boron networks is established in relation to electron transfer from uranium onto boron.

## 2- Computational method

Within DFT, the optimizations of the atomic positions and the lattice parameters were carried out using a plane wave VASP code [9, 10]. Such protocol is needed to identify minimum energy configurations of the different experimental and calculated U:B stoichiometries. The scalar relativistic calculations allowed accounting for uranium heavy elements. VASP uses the projector augmented wave (PAW) method [10, 11]. DFT exchange-correlation XC effects were approximated using the generalized gradient approximation (GGA) [12]. For each uranium boride, parameter-free unconstrained calculations were carried out with an increasingly high integration of the Brillouin zone in order to reach the ground state configuration. The individual atomic energies of U and B are deducted from the total energies to get the cohesive energy averaged per atom (cf. Table 1). The converged results allow analyses of the charge density for the purpose of extracting the magnitudes of charge transferred from U to B thanks to the approach by R. Bader in the framework of the Atoms in Molecules AIM theory [13].

Calculations were carried out assuming nonmagnetic configurations (NSP: non-spin-polarized). Further calculations accounting for two spin populations, i.e. spin-polarized (SP) lead to a magnetic polarization and long-range magnetic order only for UB$_3$ and UB$_6$ in their calculated ground state, i.e. hexagonal and cubic respectively. Their energy-volume curves in SP and NSP configurations were fitted with 3$^{rd}$ order Birch equation of states (EOS) [14].

### 3. Geometry optimization and energy-dependent results.

In so far that the dimensionality parameter highlights this work, the search of new U:B binary compounds goes through considering for each stoichiometry both 2D and 3D boron network structures, i.e. by proposing besides the experimental structures of $UB_2$ [3], $UB_4$ [15] and $UB_{12}$ [16], model ones. The purpose is to identify an energy crossover upon going from 2D to 3D boron network-based structures and to establish a connection with the amount of charge transfer onto B. Consequently, as illustrated in Fig. 1 and described in Table 1, the model structures are $UC_2$–type tetragonal structure for $UB_2$; 2D and 3D structures for hypothetic $UB_3$ –needed to complete the series– were based on $LaNi_5$-type and α-$BiF_3$-type respectively; hexagonal $UB_4$ was considered with uranium sandwiched between two *B2* layers, $U_2B_6$ as well as $UB_6$ and $UB_{12}$ in extended honeycomb *B6* and *B12* networks (cf. [17] and therein cited refs). Note that other 2D and 3D model structure candidates were considered and calculated showing trends similar to the results in Table 1.

Fully unconstrained geometry optimizations with successive iterations cycles and increasing precision integration of the Brillouin zone were carried out for all U:B chemical systems. For the experimentally known binary compounds, the results presented in Table 1 show calculated parameters in agreement with the experimental values –between brackets– Nevertheless, a trend to smaller *a* and *c* lattice constants is observed. Regarding atomic positions, small deviations versus experiments can be found. This stands particularly for $UB_4$ and $UB_{12}$ which have general positions. A general trend is the lager d(U-B) versus d(B-B) distance; the latter varies with the dimensionality and the compositions but the magnitudes remain within 1.6 – 1.8 Å range. The d(U-B) magnitudes vary around the sum of the atomic radii, r(U)= 1.75 Å and r(B)= 0.85 Å, pointing out to covalent bonds.

The discrimination between the two structural symmetry hypotheses: 2D v/s 3D for the B substructure is readily done from the total energies in so far that one formula unit (FU) is explicitly accounted for, except for experimental $UB_4$ which has 4 FU /cell in both; then $E_{Tot}$= -40.5 eV/FU. Fig. 2 shows the plots of the 2D and 3D total energies as per 1 FU. For $UB_2$ and $UB_3$, the hexagonal structure energy is lower than the cubic ones; but the opposite trend is observed for the other phase, $UB_4$, $UB_6$, and $UB_{12}$. Specifically, $UB_6$ is confirmed in the cubic $CaB_6$ structure as proposed *ad hoc* back in 2000 [8] and in agreement with $ThB_6$ structure parameters [18]. The 2D/3D crossover is just after $UB_3$. Also upon visual inspection, one can notice that the difference in 2D-3D energy increases upon the increase of boron content (see the vertical spacing between black square and red dots).

The charge transferred onto boron, δ-, is plotted against composition in the second panel of Fig. 2 where the crossover region is indicated. The decrease in magnitude (lesser charge transferred) varies linearly up to $UB_6$ and tends to flatten at ~0.1 for $UB_{12}$. Then a |δ| magnitude close to |0.46| is needed to preserve a structure with 2D boron network. Regarding this point, further checking with Th and Y borides lets confirm this result.

The cohesive energies averaged per atom illustrate further the observations above and let predict the most stable variety for each composition. Regarding the extension of the boron substructure, we considered a "double $UB_3$", i.e. $U_2B_6$ based on the proposition made for $Cr_2B_6$ in a recent work [5]. The calculated parameters and the crystal structures are provided in Table 1 and Fig. 1 respectively. Comparison with $UB_3$ regarding energy from atom averaged cohesive energy favors the $U_2B_6$ structure by -0.8 eV which, we assume, is due to the extension of the boron network from 3 to 6. However, from the calculations (Table 1) such a hexagonal layer *B6* network is not viable for $UB_6$ which adopts a cubic $CaB_6$-type structure, like ThB6 [18].

At this point in the analysis of the different results, one arrives in a relationship:
*charge transfer amount on B ⇔ dimensionality of the boron network*
with the trend that when less than 0.4é are transferred, a 2D boron network is no more 'viable' and it becomes 3D.

## 4      Analyses of the magnetic states of $UB_3$ and $UB_6$

*Energy-volume equations of states and magnetic states*

A remarkable feature pertaining to the ground state of most $UB_n$ under study –i.e., except for $U_2B_6$ where d(U-U) 3Å is that two successive U atoms are separated by a distance equivalent to a lattice parameter, *a* or *c* parameter, i.e. ~3.6 Å which happens to be slightly larger than the Hill critical distance of 3.5 Å [19]. Such separation is needed to allow the onset of magnetic polarization on U-5f, i.e. below this distance intra-band spin polarization is not allowed because U-5f bands start overlapping [20]. This feature incited further calculation assuming two spin channels, in spin-polarized SP configuration, starting from the converged NSP (non-spin-polarized) calculations.

However, the Hill criterion is only indicative and accurate SP calculations are needed.
Indeed, at self-consistent convergence, stable magnetic solutions were found only for hexagonal $UB_3$ and cubic $UB_6$. The spin-only magnetizations of $1.47\mu_B$ and $2.40\mu_B$ (Bohr magnetons) are accompanied by a lowering of the total energy while little changes of the structure could be identified.

In order to have a complete view of the volume change of energy, we carried out (E, V) sets of calculations around minima found from the geometry optimization for both SP and NSP configurations. The fit of the E-V curves from the calculations around minima is done based on the equation of state EOS by Birch [14]:

$$E(V) = E_o(V_o) + [9/8]V_o B_o[((V_o)/V)^{[2/3]} - 1]^2 + [9/16]B_o(B' - 4)V_o[((V_o)/V)^{[2/3]} - 1]^3,$$

where $E_o$, $V_o$, $B_o$ and $B'$ are the equilibrium energy, the volume, the bulk modulus, and its pressure derivative. In both cases calculated $B' = 3.6$.

Figure 6 displays the SP and NSP E-V curves and the resulting fit values with 3$^{rd}$ order EOS. In both $UB_3$ and $UB_6$, the SP curve is below the NSP one at larger volumes (ascending branch on the right-hand side) with a trend to merge together at low volumes (branch on the left-hand side). The energy difference SP-NSP is small -0.06 eV for $UB_3$ but three-times larger: -0.20 eV for $UB_6$. The zero pressure bulk modulus $B_0$ is higher in the NSP configuration because of the smaller volume; it is found in the range of 200 GPa. Such low magnitude, ex. versus diamond for instance which is characterized by $B_0 \sim 400$ GPa, has different origins according to the structure and its openness i.e. layered $UB_3$ and open (noncompact) $UB_6$ with large U-B separation.

In so far that spin-polarized calculations assume implicitly a ferromagnetic order, further antiferromagnetic calculations were carried out assuming double cells with two magnetic subcells, one oriented as SPIN UP, the other as SPIN DOWN. The energy was then found to increase with a lowering of the magnetization magnitude down to 0.67 $\mu_B$ for $UB_3$ and 1.71 $\mu_B$ for $UB_6$. Then one can assume that the ground state in both compounds can be considered as ferromagnetic. Lastly, we point out that spin-only present results should be completed with spin-orbit coupling LS calculations needed to account for 5f bands in order to comply with Hund's 3$^{rd}$ –cf. [21] and therein references.

The relevant heavier calculations are underway in collaboration with European universities.

The illustration of the magnetic charge density resulting from the difference between spin ↑ charge and spin ↓ is shown in Fig. 4 for the two magnetic borides. The difference in color and thickness of the shells on uranium spheres is due to the magnitudes of the magnetic moments carried by uranium in the two compounds, i.e. much larger in UB$_6$ than in UB$_3$. Also, the absence of any magnetic charge density on boron clearly shows that there is no induced moment on B which would arise from the U-B bonding. This signals to a certain extent the isolated character of uranium valence states, larger in UB$_6$ with d(U-B)~2.96 Å and highlights the subsequent onset of intraband spin polarization.

Further to these observations, the volume change of magnetization is shown in Fig. 5 where V$_0$ indicates the equilibrium volume (cf. EOS parameters in Fig. 3). Both curves show an 'S' like behavior but the switching from a nonmagnetic state to a magnetic one is clearly different in the two compounds: a smooth evolution versus an abrupt volume change of the magnetization respectively in UB$_3$ and UB$_6$.

*Electronic density of states of UB$_3$ and UB$_6$*

The site and spin projected density of states (DOS) were subsequently calculated for NSP and SP configurations using the corresponding calculated lattice parameters of Table 1. Fig. 6 shows the site projected DOS for NSP (left-hand-side LHS) and SP (right-hand-side RHS) calculations. The zero of energy along the *x*-axis is with respect to the Fermi level (E$_F$) which is crossed by a large uranium 5f density of states in both borides; note that the mainly empty U-5f band is centered above E$_F$ within the conduction band. The boron states with s,p characters are smeared throughout the valence band VB. Such large DOS at E$_F$ signals an instability of the nonmagnetic configuration, which is lifted through calculations assuming two spin channels: majority spin ↑ and minority ↓. Indeed the RHS panels show the down energy shift of ↑-DOS and up energy shift of ↓, larger in UB$_6$. The difference of populations provides the magnetic moment which is smaller in UB$_3$ than in UB$_6$ as discussed above.

## 4    Concluding remarks.

Our revisit of the U:B system has provided results that allowed casting a quantitative assessment of the crystal symmetry changes upon increasing the amount of boron in a discrete manner in $UB_n$. Specifically, by introducing new compositions, $UB_3$, $U_2B_6$, and $UB_6$, examined together with experimentally identified chemical systems $UB_2$, $UB_4$ and $UB_{12}$, all were calculated and structurally and energetically assessed within DFT in 2D and 3D boron substructure symmetry. A structural crossover 2D→3D was then identified in relation to the amount of charge departing from U towards B (~0.46). Energy volume EOS's allowed identifying two (Ferro)magnetically stable compounds: $UB_3$ and $UB_6$. With opposite volume change of the magnetization, respectively soft and abrupt.

# Table 1 in 3 parts

Finite stoichiometries in the U:B system. Experimental and "hypothetic" binary borides $UB_n$. Lattice constants and distances are expressed in the unit of Å. Energies are in units of eV. For actual borides, the experimental data are between brackets. Calculated atomic constituent's energies: E(B)= -5.33 eV. E(U)= -11.01 eV. "mult." Indicates the multiplicity of the relevant Wyckoff position.

| Part 1 | $UB_2$ | | "$UB_3$" | | "$U_2B_6$" |
|---|---|---|---|---|---|
| Space group | P6/mmm | I4/mmm | $LaNi_5$-derived P6/mmm | α-$BiF_3$-type Cubic. Fm-3m | P6/mmm |
| a<br>c | 3.05 (3.13)<br>4.04 (3.99) | 3.58<br>6.44 | 3.59<br>3.86 | 5.52 | 5.21<br>3.94 |
| Atomic positions | U: 0,0,0<br>B: 1/3, 2/3, ½ (mult.2) | U: 0,0,0<br>B: 0,0,0.376 (mult.2) | U: 0,0,0<br>B: ½, 0, ½ (mult.3) | U: 0,0,0<br>B1: ½, ½, ½<br>B2: ¼ ¼ ¼ (mult.2) | U: 1/3, 2/3, ½ (2fold)<br>B: 0.332, x, 0 (mult.6) |
| d(U-B) | 2.68 (2.69) | 2.42 | 2.64 | 2.76 | 2.63 |
| d(B-B) | 1.76 (1.80) | 1.59 | 1.80 | 2.38 | 1.73 |
| Tot. Energy | -26.37 | -23.55 | -31.14 | -28.57 | -63.72 |
| Coh. energy /at. | -1.57 | -0.63 | -1.04 | -0.33 | -1.22 |
| *Bader charge* $B^{-\delta}$ | -0.51 | -0.77 | -0.46 | -0.25 | -0.43 |

| Part 2 | $UB_4$ | |
|---|---|---|
| Space group | Experimental P4/mbm | P-6m2 N°187 |
| a<br>c | 7.02 (7.08)<br>3.94 (3.98) | 3.03<br>6.10 |
| Atomic positions | *U: 0.31,0.81,0 (idem)<br>*B1: 0, 0, 0.2(0.21)<br>*B2: -0.087(0.086), 0.413(0.4144), ½ (*mult.4)<br>B3: 0.179(0.174), 0.039(0.040) ½ (mult.8) | U: 0,0,0<br>B1: 1/3, 2/3, 0.182 (mult.2)<br>B2: 0, 0, 0.136 (mult.2) |
| d(U-B) | 2.67 – 2.70 | 2.63 |
| d(B-B) | 1.58 – 1.75 | 1.69-1.77 |
| Tot. Energy | -162.0 | -36.2 |
| Coh. energy /at. | -1.64 | -0.79 |
| *Bader charge* $B^{-\delta}$ | -0.39 | -0.21 |

| Part 3 | "UB$_6$" | | UB$_{12}$ | |
|---|---|---|---|---|
| Space group | *Pm-3m* | *P6/mmm* | *Fm-3m* | *P6/mmm* |
| *a*<br>*c* | 4.034 | 5.20<br>3.94 | 7.433 (7.468) | 5.14<br>5.78 |
| Atomic positions | U: 0,0,0<br>B: ½ ,½, 0.197<br>*(mult.6)* | U: 0,0,0<br>B:0.332,0,½,<br>*(mult.6)* | U(0,0,0) B: ½, ½ 0.169(0.161),<br>*(mult12)* | U: 0,0,0<br>B:0.339,0.339, 0.281<br>*(mult.12)* |
| d(U-B) | 2.96 | 2.62 | 2.78 | 2.38 |
| d(B-B) | 1.73 | 1.74 | 1.77 | 1.65 |
| Total Energy | -53.08 | -48.56 | -90.76 | -83.62 |
| Coh.Energy /at. | -1.44 | -0.79 | -1.28 | -0.66 |
| *Bader charge B$^{-\delta}$* | -0.31 | -0.28 | -0.15 | -0.11 |

# **F I G U R E S**

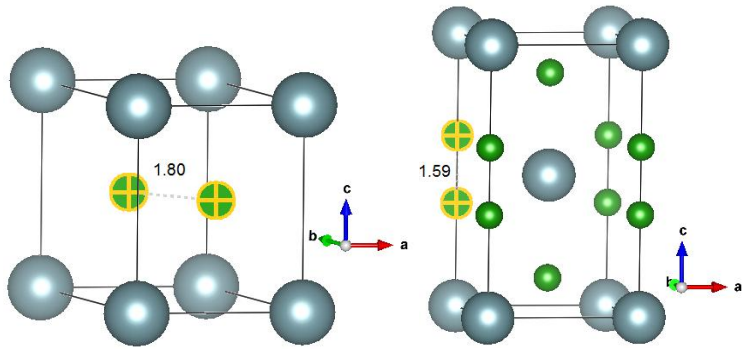

hexagonal          tetragonal

**$UB_2$**

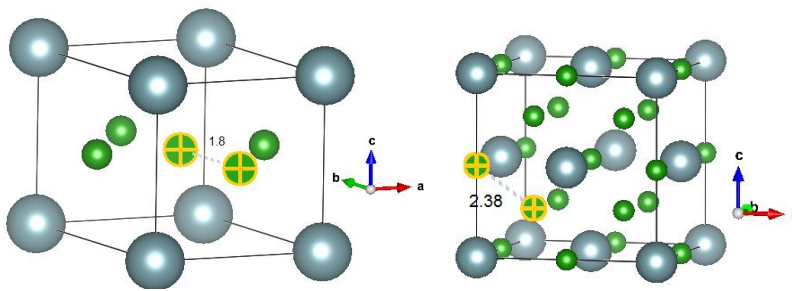

**hexagonal**          **cubic**

**"$UB_3$"**

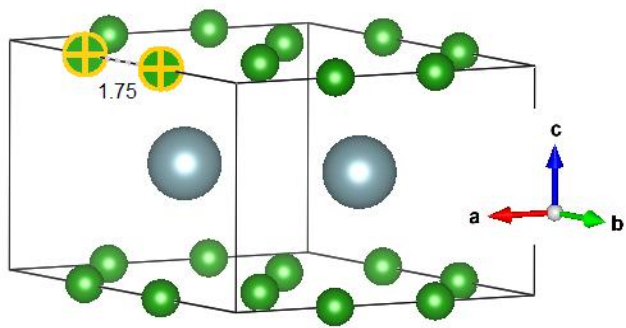

**"$U_2B_6$" hexagonal**

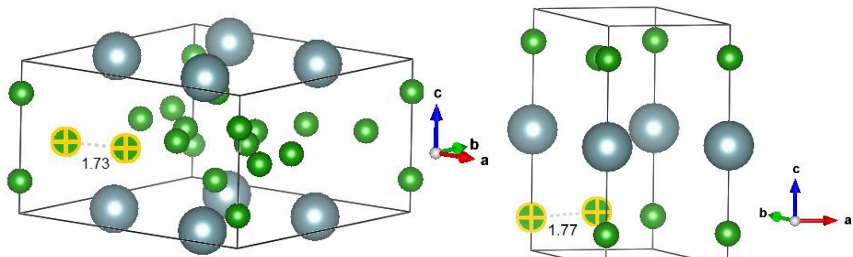

tetragonal          "hexagonal"

**$UB_4$**

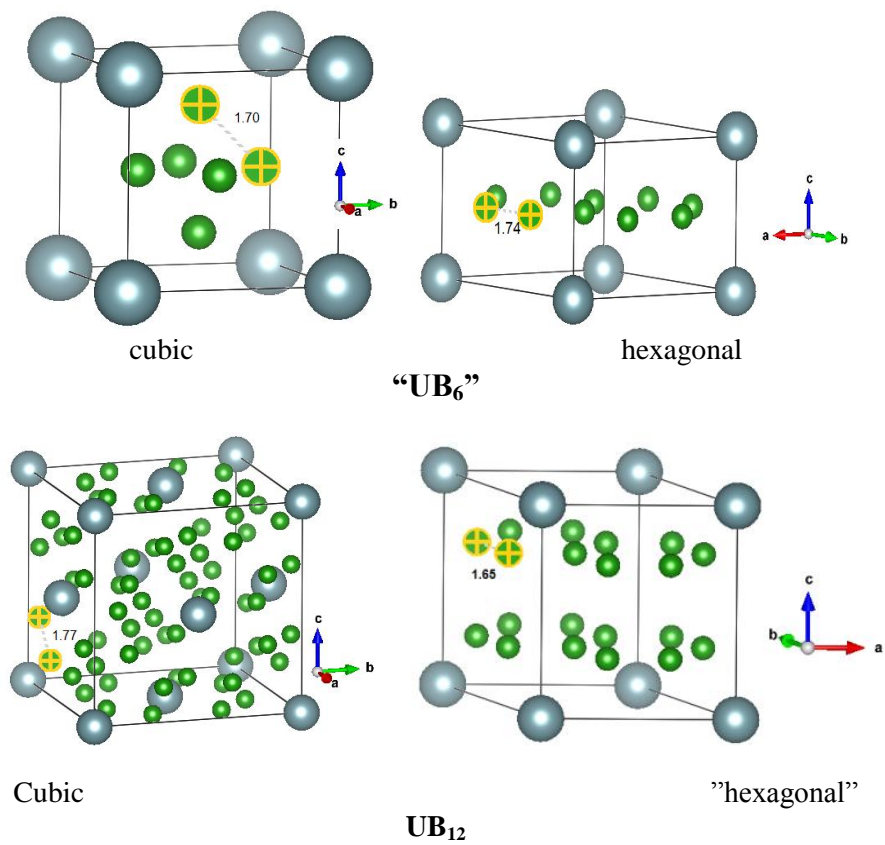

Figure 1. Experimental and "hypothetic" compounds proposed as completing the U:B system.

The B-B closest distances are shown.

Large grey and small green spheres correspond to U and B respectively

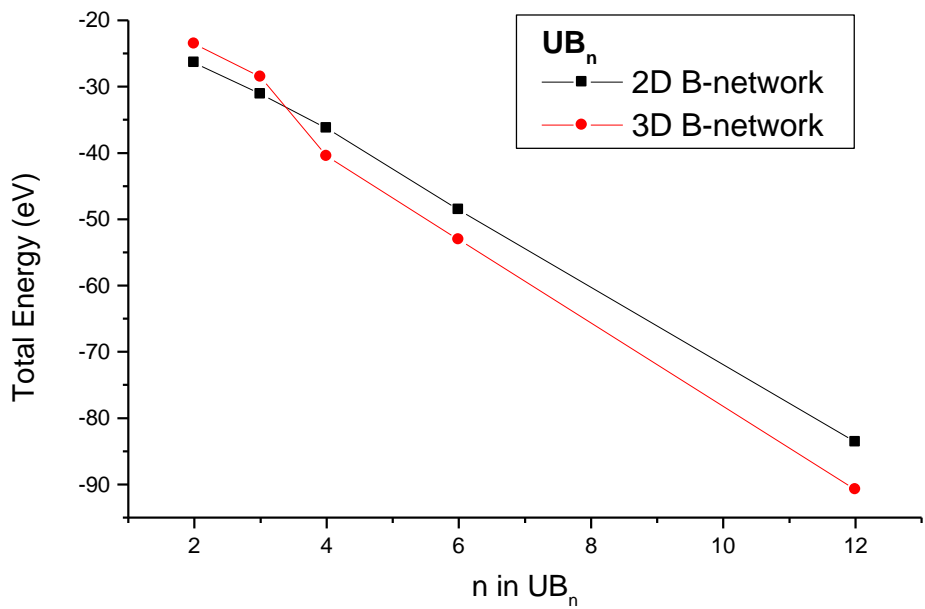

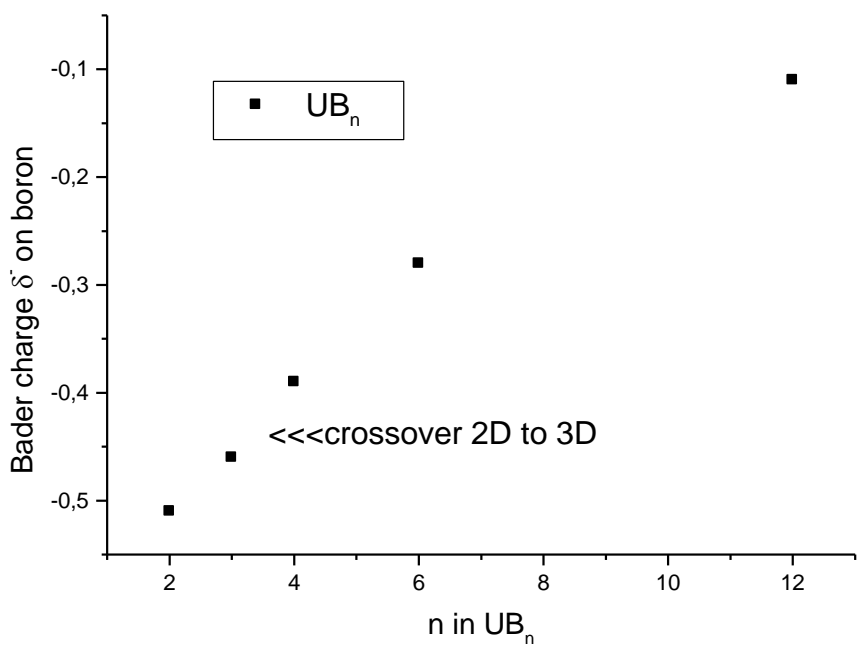

Figure 2. 2D/3D lattice symmetry crossover.

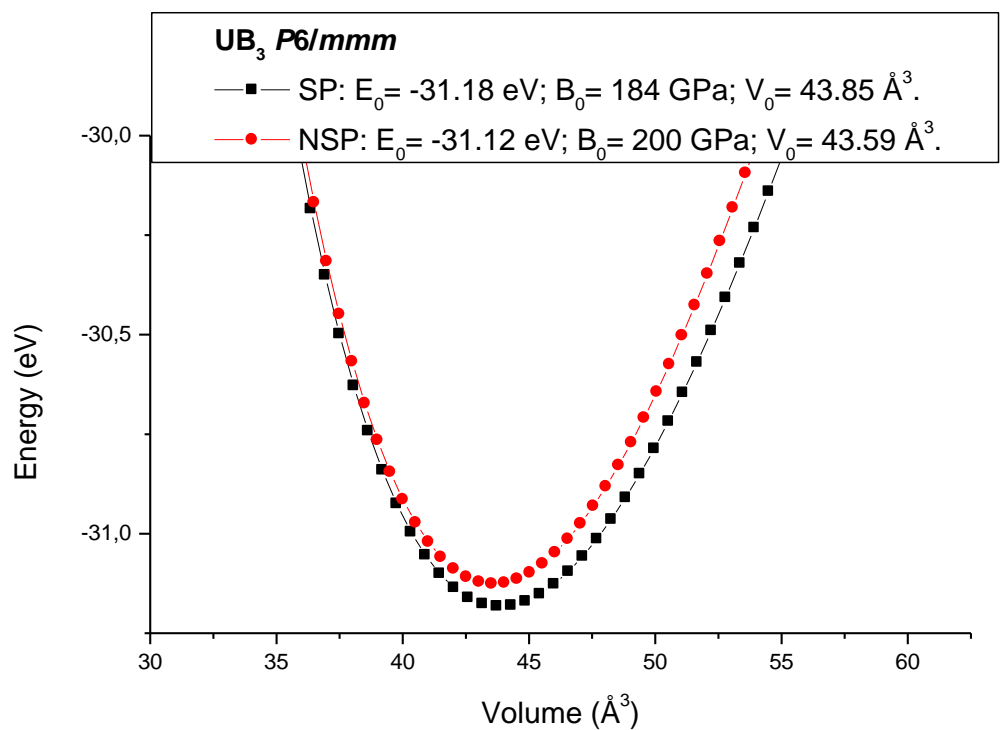

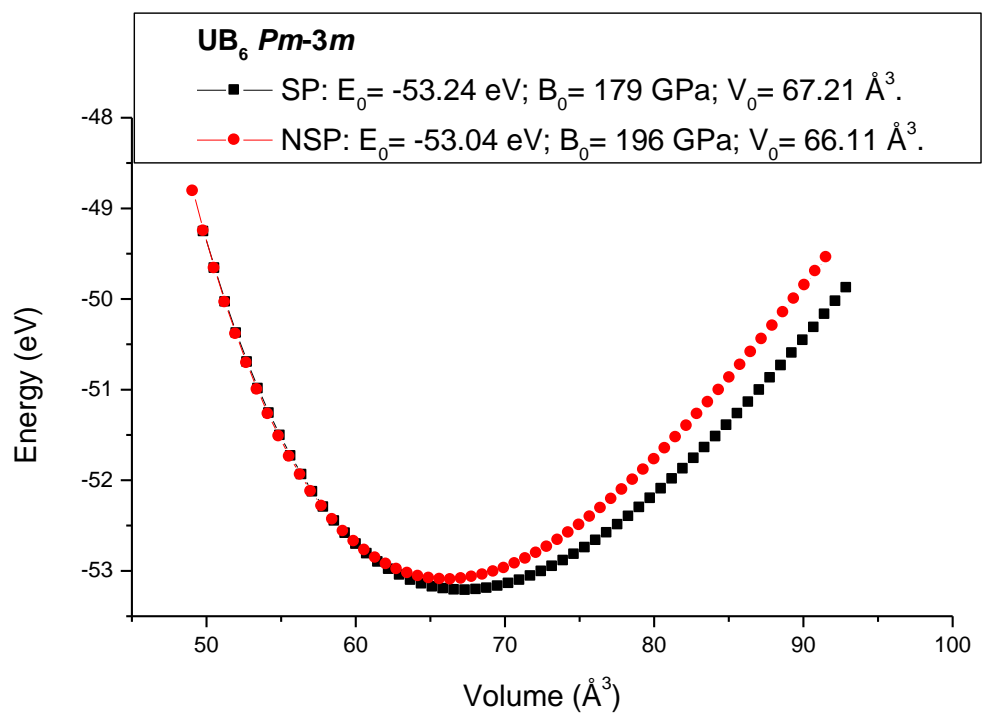

Figure 3: Energy-volume curves and fit values from Birch 3$^{rd}$ order equation of state (EOS).

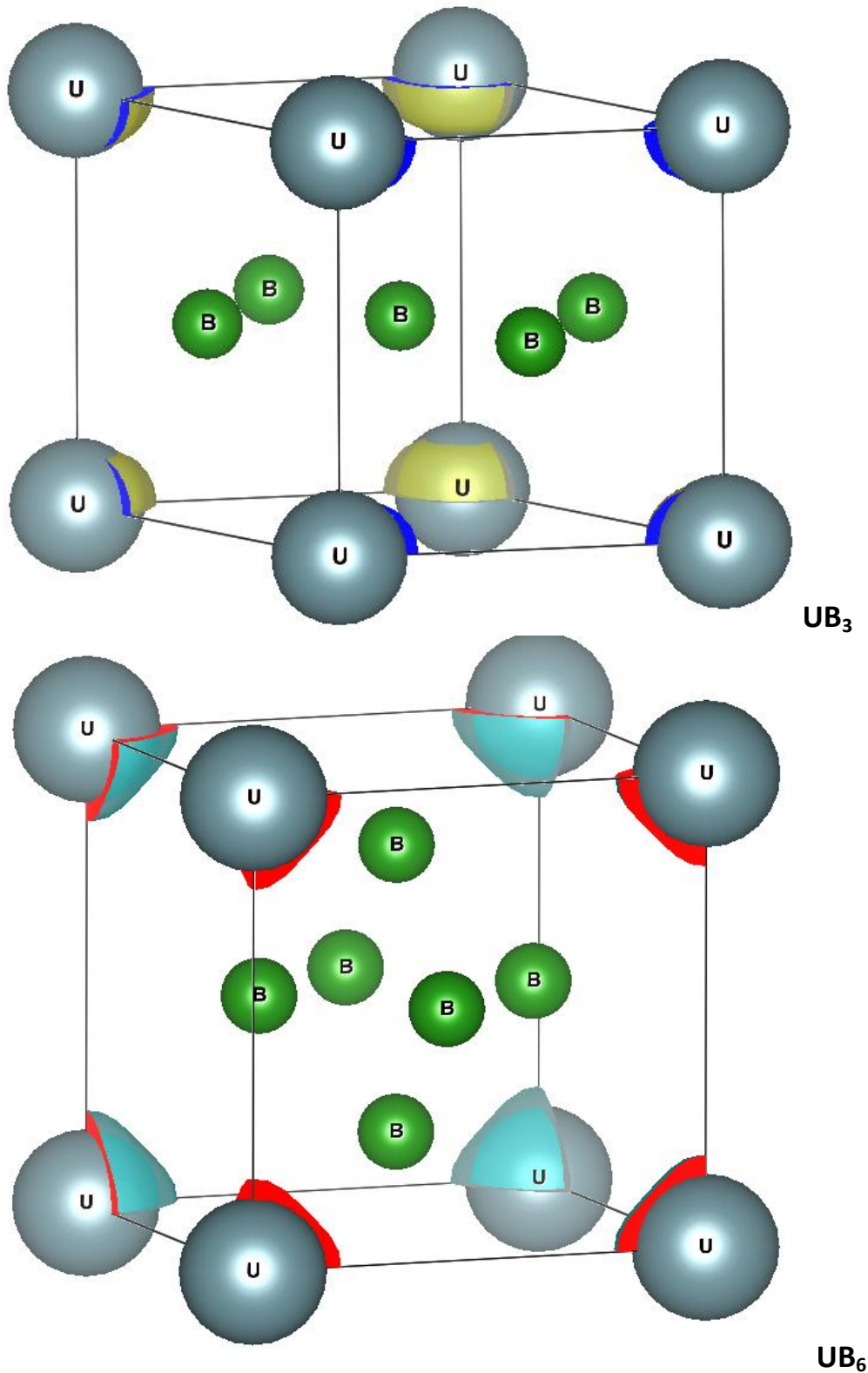

Figure 4. Magnetic charge density envelops on uranium in UB$_3$ and UB$_6$ identified as ferromagnetic in the ground state with 1.47$\mu_B$ /FU and 2.40$\mu_B$ /FU (Bohr magnetons per Formula Unit).

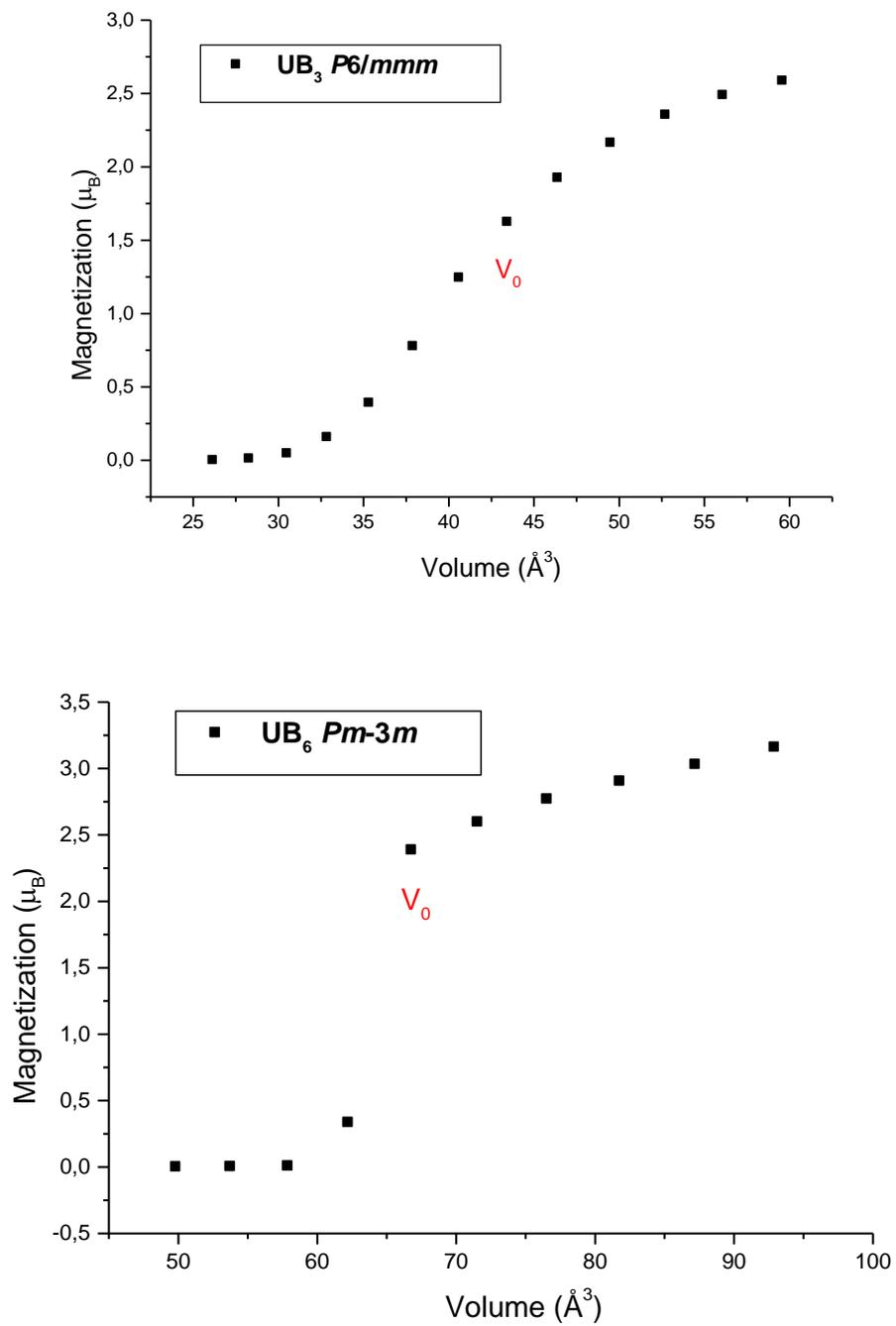

Figure 5. Smooth and abrupt volume change of the magnetization respectively in $UB_3$ and $UB_6$. $V_0$ indicates the equilibrium volume (cf. Fig. 3).

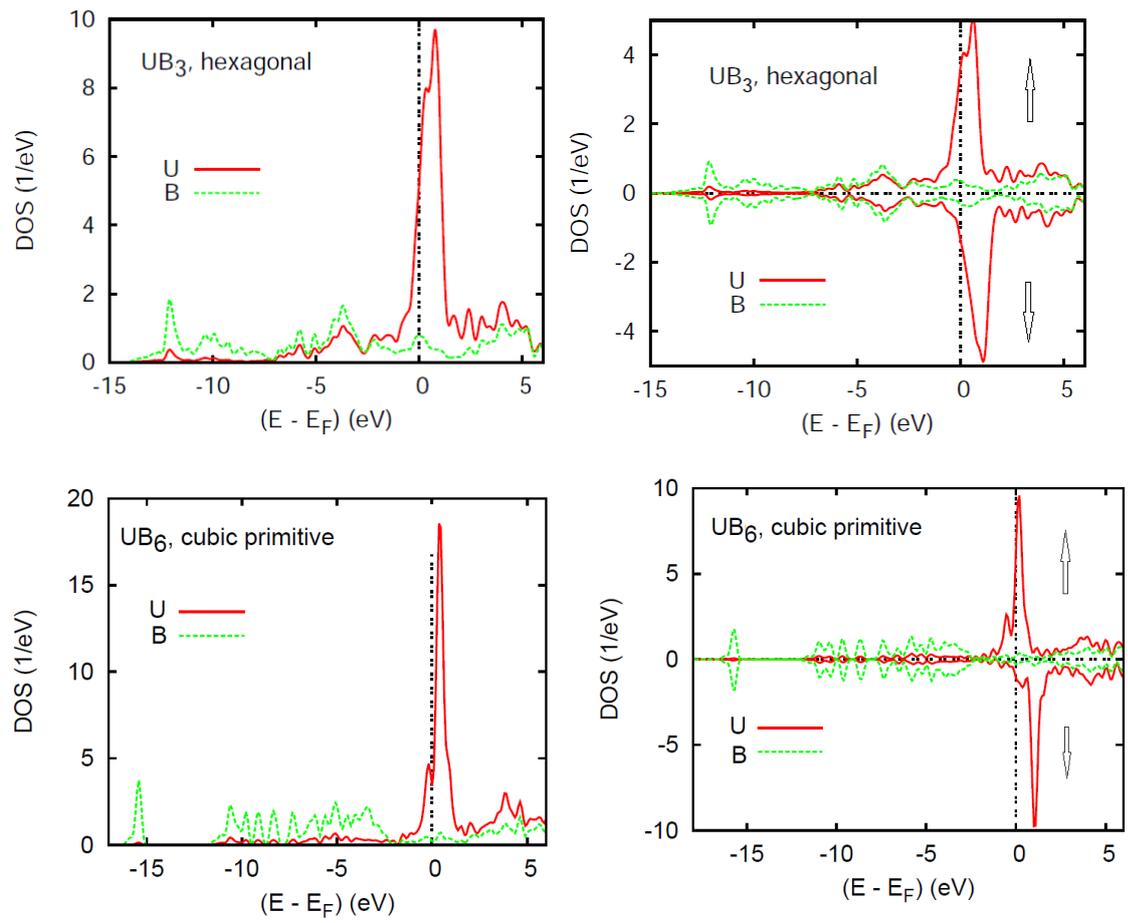

Figure 6. NSP (LHS) and SP (RHS) electronic density of states of UB$_3$ (top) and UB$_6$ (bottom).